\documentclass{article}

\usepackage{latexsym}

\begin{document}

\centerline{\Large{An extended phase space for Quantum Mechanics}}

\vskip 0.5 truecm

\centerline{C. Lopez}

\vskip 0.3 truecm

\centerline{\small{Dpto. de F\'{\i}sica y Matem\'aticas, Fac. de Ciencias}}
\centerline{\small{UAH, E-28871 Alcal\'a de Henares (Madrid), SPAIN}}
\centerline{\small{carlos.lopez@uah.es }}

\vskip 0.3 truecm

\begin{abstract}

The standard formulation of Quantum Mechanics violates locality of interactions and the action reaction principle. An alternative formulation in an extended phase space could preserve both principles, but Bell's theorems show that a distribution of probability in a space of local variables can not reproduce the quantum correlations. 

An extended phase space is defined in an alternative formulation of Quantum Mechanics. Quantum states are represented by a complex va\-lued distribution of amplitude, so that Bell's theorems do not apply.

\vskip 0.3 truecm

\end{abstract}	

\vskip 0.5 truecm

\section{Introduction}

According to the EPR argument\cite{EPR}, the standard formulation of Quantum Mechanics (QM) is incomplete. The authors did not consider the possibility that measurements could have non local effects. On the other hand, Bell's theorems \cite{Bell} prove that probabilistic theories with  local hidden variables  contradict the quantum (and experimental \cite{Aspect}) correlations. 

The action reaction principle (ARP) is a fundamental ingredient of Me\-cha\-nics. In its simplest formulation, it states that
when two systems interact both of them depart from their isolated evolution law. Otherwise, there would be changes of state in a system without any cause. 

In section 2, a simple argument shows that the ARP is violated in the standard formulation of QM. If it happens in the framework of Classical Mechanics the phase space is incomplete, and additional variables of state will restore the ARP. The same hypothesis, incompleteness of QM, is considered in this paper.

Section 3 presents another paradox between the ARP and indirect measurements on an elementary particle,
when some virtual paths are discarded. If interactions are local, we conclude that there is an accompanying system, the de Broglie wave\cite{deBroglie}. Additionally, there must be variables of state for the corpuscular component, the point particle. The quantum particle is a composite.

The thesis of Bell's theorems can be avoided if some hypothesis is not fulfilled by the mathematical model. A quantum state will be represented, as in standard QM, by a complex valued distribution of amplitude in a space of state variables. In section 4 a generic quantum system, with an arbitrary family of magnitudes as variables of state, is considered, generalizing the case of spin variables presented in \cite{mio}.  Relative frequencies are obtained by application of Born's rule\cite{Born} to the marginal amplitudes, so that the typical interference phenomenon in the superposition appears. Complex amplitudes, as (real instead of non negative) quasi probabilities\cite{Wigner}, do not fulfil the hypothesis of Bell's theorems, which do not apply to this formalism.

The phase space of three particular systems is described in the following sections: isolated spinless point particle, the paradigmatic two slit experiment, and spin variables, both for an individual particle and for a composite of two in a total null spin state (Bell's experiment). Predictions of standard QM are reproduced. 

The proposed formalism is just a new interpretation of QM, up to the wave like accompanying system\footnote{This hypothesis was considered even before the establishment of the standard formalism.} and its observable effects, when it is spatially isolated from the corpuscular component. Vacuum must be the background of the wave like system. Last section contains some remarks about its physical interpretation.

\section{The action reaction principle in QM}

The following analysis shows that the standard formulation of QM either is incomplete or it violates the ARP. 

Let $A$ be a self-adjoint operator, representing a physical magnitude of a quantum system in the associated Hilbert space, and 
$|a_1>$ an eigenvector of $A$ with eigenvalue $a_1$, representing the initial state of the system. A measurement of magnitude $A$ is performed; the result of measurement is $a_1$ and the final state $|a_1>$, trivial projection of state. The pointer has moved from ``neutral'' initial position to ``$a_1$''. One of the systems in interaction (the apparatus) has changed of state and the other has not. The action reaction principle (ARP) is violated. The ARP is violated in standard QM because the maximal number of commuting operators is lower than the dimension of the classical phase space.\footnote{This paradox also appears in the Hamiltonian evolution, and not exclusively under measurement. Let $Y$ be a magnitude of another system, 
and $H(A,Y)= AY$ the Hamiltonian of interaction. The initial state of the second system is $|d>$ $=\sum _k z_k|y_k>$, eigenstate of magnitude $D$, $[Y,D]\neq0$. From the initial state of the composite, $|S>(0)=$ $|a_1>|d>$, and the Hamiltonian $AY$ we get $|S>(t)=$ $|a_1>$ $(\sum z_k(t)|y_k>)$, with $z_k(t)=$ $exp(i/\hbar$ $a_1y_k t)z_k$.
Magnitude $D$ of the second system evolves while the first system remains in its initial state.}

For another arbitrary magnitude $B$, $[A,B]\neq 0$, we can express the initial state as 
$|a_1> = \sum _j z_j|b_j>$. The system in state $|a_1>$ does not have definite value of $B$. If $B$ instead of $A$ were measured, some value, say $b_1$, would be obtained. Let us consider the  hypothesis that $b_1$ is a hidden value in the quantum state; being $B$ arbitrary, the same hypothesis applies to all magnitudes. An extended phase space contains variables giving account of the  result of arbitrary hypothetical measurements. When $A$ is measured, some of these variables change of value and the ARP is preserved.

We describe quantum states through complex valued distributions of amplitude in a space of state variables. In the
space of $N$-tuples of values of $N$ physical magnitudes 
${\cal P} \equiv \{(a_i,b_j,c_k,\ldots)\}$,
the orthodox quantum formalism assigns a family of distributions of amplitude 
$\sum _i z_i|a_i>$\footnote{$z_i = \delta _{i1}$ in the particular case
$|a_1>$ previously considered.}, $\sum _j z_j|b_j>$, $\sum _k z_k|c_k>$, etc., on each space  $\{a_i\}$, $\{b_j\}$, $\{c_k\}$, \ldots respectively,  sets of eigenvalues of  each magnitude. The correspondence between these distributions is determined by the change of bases.

We will consider an alternative formulation, a distribution $Z_{ijkl\cdots}$ in 
$\cal P$ whose marginals match those of standard QM, i.e., $\sum _{jkl\cdots} Z_{ijkl\cdots} :: $ $z_i$,
$\sum _{ikl\cdots} Z_{ijkl\cdots} :: $ $z_j$, $\sum _{ijl\cdots} Z_{ijkl\cdots} :: $ $z_k$, etc., where $::$ means projective equivalence.
The distribution $Z_{ijkl\cdots}$ in $\cal P$ will reproduce the correspondence between standard QM distributions,
that is, $z_j \leftarrow Z_{ijkl\cdots} \to z_k$ through marginals will replace
 $z_j \leftrightarrow z_k$ through change of bases.

If $\lambda _1 = (a_1,b_1,c_1,\ldots)$ were a complete description of state of an individual system, we would expect 
the existence of a classical distribution of probability $P(\lambda )$ in $\cal P$, representing an ensemble of independent systems. 
Bell's theorems show that $P(\lambda )$ does not exist. Wave particle duality and the two slit experiment suggest that there is an additional, accompanying wave like system, the de Broglie wave.

\section{Elementary particle as a composite}

Let $z_1|a_1> + z_2|a_2>$ be the vector of state of an elementary particle, such that
$|a_1>$ and $|a_2>$ components have spatially separated wave packet representations, i.e., follow different virtual paths\footnote{For example,
a spin $1/2$ particle going through a Stern-Gerlach apparatus.}. 
A particle detector is placed in one of the paths, say $|a_1>$. There is not detection (indirect, negative measurement). The final state of the system is $|a_2>$, different from the initial one. The detector has not (apparently) changed of state, and the ARP is violated.

There are possibly unobserved variables in the detector that change of value along the process. After all, the detector is designed to show an observable reaction when a particle hits it, i.e., through a local interaction. Perhaps the particle following the other path interacts with the detector in an unknown and non local way.

Let us suppose that interactions are local. The particle following path $|a_2>$ does not interact with the spatially separated detector, located at path $|a_1>$.
There must be another physical system, the de Broglie wave, that locally interacts with the detector, because there is a change of state in the (composite) system caused by the detector. Perhaps we could find an observable reaction in an appropriately designed detector, looking for a wave like system \footnote{This effect is not described in standard QM, but does not contradict it, as far as magnitudes and relative frequencies for direct measurements of the particle are preserved.}.

We will suppose that an isolated elementary particle is
a composite of corpuscular and wave like subsystems. In the previous experiment, the corpuscular subsystem has a definite position coordinate (the spatial wave packet $|a_2>$),
while the wave subsystem has two components at $|a_1>$ and $|a_2>$.
The distribution of amplitude represents an ensemble of physical systems,
and a statistical correlation between wave and corpuscular subsystems through Born's rule.
Notice that the negative result of measurement is not just a reset of information about the state of system, because there is an observable 
change of state \footnote{If the initial state is spin up in direction $X$, and the particle goes through two
Stern--Gerlach systems in directions $Y$ and $-Y$, the final (reconstructed) state is again spin up in direction $X$, which can be 
measured. A detector is now located at the $Y$ spin down path. 
The particle is not detected, so that it has $Y$ spin up state before a final $X$ measurement. In a particular case, the final measurement in direction $X$ gives spin down result, against the behaviour in absence of detector. The presence of the detector has changed the final state of this particle.}.

In the alternative formulation sketched in the previous section,  $\lambda _1 =$ $(a_1,b_1,$ $c_1,\ldots)$ is not a complete
description of state of an individual system. In the former example the particle has precise value of the position magnitude,  $x_2$ at the wave packet  
$|a_2>$,  and does not interact (locally) with the detector. Instead, the wave component at $|a_1>$ does, and later on the interaction between a perturbed wave and the particle gives account of a new final state of the particle.
The correlation (interaction) between particle and wave can not be represented by a distribution of probability on the space of state variables; interference phenomena appear in the superposition of complex amplitudes, but not with classical probabilities.

\section{Extended phase space}

In Classical Mechanics the phase space of a system is usually finite dimensional and physical magnitudes are real functions on it; there
are functional relations between them, as energy and angular momentum depending on position and momentum, etc.
In standard QM, magnitudes are represented by non commuting operators,  and can not generically  have  precise values jointly, common eigenvectors. The classical functional dependence between magnitudes is not fulfilled by the quantum eigenvalues\footnote{For example, for spin operators $S_{\theta} =$ $\cos (\theta) S_x +$  $\sin (\theta) S_y$ the corresponding expression for eigenvalues is wrong.
However, for a magnitude $F$ functionally dependent on two or more commuting ones,
$F=f(A,B)$, $[A,B]=0$, the same functional dependence is fulfilled by the eigenvalues, 
$f_{ij} = f(a_i,b_j)$.}.

The extended phase space of a quantum system is defined in two steps. First, a set $\cal P$ of $N$-tuples of all possible (eigen)values of $N$ physical magnitudes. Magnitudes functionally dependent of two or more commuting ones are redundant, while those functionally dependent of non commuting magnitudes are not. All functions of the classical phase space could be in the list, modulo redundancies, for a maximal resolution\footnote{The set $\cal P$ is therefore infinite dimensional. However, for practical purposes a finite number of magnitudes is enough.}. In a second step, a distribution of amplitude is defined in this set $\cal P$. 

Let $\cal H$ be the Hilbert space of a quantum system in orthodox QM, 
and ${\cal F} =\{A,B,C,\ldots\}$ a family of self adjoint operators representing physical magnitudes of the system. ${\cal A} =\{a_i\}$, ${\cal B} =\{b_j\}$,
${\cal C} =\{c_k\}$, \ldots, are the sets of eigenvalues of each corresponding operator, possible values of the magnitude. Then 

\begin{equation}
{\cal P} = {\cal A} \times {\cal B} \times 
{\cal C} \times {\cal D} \times \cdots \quad \lambda = 
(a_i,b_j,c_k,d_l, \ldots ) \in {\cal P}
\end{equation} 
is the set of $N$-tuples of values of these magnitudes.
A distribution of amplitude in $\cal P$ is a complex function 
$Z(a_i,b_j,c_k, \ldots)$ = $Z(\lambda)$.

A state of the physical system is represented by a pair $[\lambda _0, Z]$,  
$\lambda _0 \in {\cal P}$.
The first component $\lambda _0$ determines the value of all magnitudes in this state. $\lambda _0$ contains therefore the result of every hypothetical direct measurement of an arbitrary magnitude. $Z$ represents the composite\footnote{An ensemble 
of individual particle plus wave systems, where 
$\lambda _0$ takes all allowed values.},
and the (possibly stochastic) interaction between wave and particle subsystems.

Being $\lambda _0$ hidden, we can at most calculate relative frequencies for
any ma\-ximal subset of compatible magnitudes from the distribution $Z$; in fact, 
we can get relative frequencies for any subset of magnitudes, but if they are incompa\-tible the corresponding relative frequencies are just formal, not observable.  The relative frequencies are obtained in two steps, computing first the marginal amplitudes from $Z$, and then applying the usual Born's rule.
For example, relative frequencies for a joint measurement of $A$ and $B$ \footnote{$[A,B]=0$ and $\{A,B\}$ a maximal set of compatible magnitudes, i.e., $\{|a_i,b_j>\}$ a basis of the Hilbert space.} are found as follows. The marginal amplitudes are

\begin{equation}
Z(a_i,b_j) = \sum _{k l\cdots} Z(a_i,b_j,c_k,d_l \ldots)
\end{equation}
According to Born's rule, $|Z(a_i,b_j)|^2$ are proportional to the probabilities $P(a_i,$ $b_j)$\footnote{Even if we normalize $Z$, 
$\sum |Z(a_i,$ $b_j,c_k,$ $\ldots)|^2 = 1$, the associated marginals are not generically normalized.}.

Similarly, for $[B,C]=0$, we get
$P(b_j,$ $c_k)$ proportional to $|Z(b_j,$ $c_k)|^2$, $Z(b_j,$ $c_k)$ the corresponding marginals.
Even if $[A,C]\neq 0$ we can formally calculate the 
probabilities $P(a_i,b_j,c_k)$ through the marginals

\begin{equation}
Z(a_i,b_j,c_k) = \sum _{l \cdots} Z(a_i,b_j,c_k,d_l \ldots)
\end{equation}
and Born's rule, $P(a_i,b_j,c_k)$ proportional to $|Z(a_i,b_j,c_k)|^2$. 

Notice that marginal amplitudes fulfil $Z(a_i,b_j) =$ $\sum _k$ $Z(a_i,b_j,c_k)$, but the associated probabilities obtained through Born's rule don't, 
$P(a_i,b_j)$ $\neq$ $\sum _k$ $P(a_i,$ $b_j,c_k)$, because of the typical interference terms, real part of  $Z^*(a_i,$ $b_j,c_k)$ $Z(a_{i},$ $b_{j},$ $c_{k'})$. Bell's theorems apply to hypothetical $P'(a_i,$ $b_j,c_k)$ whose marginals were the observable  $P(a_i,b_j)$,
and prove that such $P'$ do not exist.

In the correspondence with the standard formalism we must take into account that a state of the system is defined by a ray in the Hilbert space. Normalization factors and arbitrary phases in the definition of a basis must be taken into account.
For example, if we get from $Z(a_i,b_j,c_k,d_l \ldots)$ the marginals $Z(a_i,b_j)$, the usual representation is a unit vector

\begin{equation}
|S> =  \sum _{i j} z(a_i,b_j) |a_i,b_j> = {\cal N} \sum _{i j} Z(a_i,b_j) \|a_i,b_j> 
\end{equation}
where $\cal N$ is the normalization factor, and $\|a_i,b_j> = exp(i\theta _{ij})|a_i,b_j>$
relates eigenvectors generated in the alternative formulation with some ``established'', used by convention, eigenvectors in the standard 
formulation\footnote{Two vectors $|S_1> = \sum _k z_k|c_k>$
and $|S_2> = \sum _k exp(i\phi _k)z_k|c_k>$ represent different states, although $P_1(c_k)=P_2(c_k)$,
because relative frequencies for some other magnitude will not match. On the other hand,
$|S_1> = \sum _k (exp(i\phi _k)z_k)[exp(-i\phi _k)|c_k>]$ is obviously the same physical state 
represented in two  orthonormal bases of eigenvectors $|c_k>$ and $\|c_k> =$ 
$exp(-i\phi _k)|c_k>$, which can be indistinctly used.}.

\section{Entanglement}

If we apply the previous formalism to a quantum system made of two or more particles, there will be
some physical magnitudes associated to individual particles (each particle position, momentum, etc.) and others associated to
the composite (potential energy, angular momentum, \ldots). 
We can define projections from the space $\cal P$ of $N$--tuples of eigenvalues of the whole family of magnitudes
onto spaces of eigenvalues of magnitudes in each subfamily, ${\cal P}^I$ for particle $I$,  
${\cal P}^{II}$ for particle $II$, \ldots,  ${\cal P}^{comp}$ for global magnitudes of the composite.
The distribution of amplitude $Z(\lambda )$, $\lambda \in {\cal P}$, can be written as $Z(\lambda ^I,$ $\lambda ^{II},$
$\ldots,$ $\lambda ^{comp})$. It is, as before, an ensemble representation of the composite,  accompanying de Broglie wave and  corpuscular components of the system.

Magnitudes of an individual particle are compatible with magnitudes of another, so that distributions of probability as 
$P(a_i^I,$ $b_j^{II})$ are observable, and are obtained through the marginals

\begin{equation}
Z(a_i^I,b_j^{II}) = \sum _{
\pi _{A^I}(\lambda ) = a_i^I; 
\pi _{B^{II }}(\lambda ) = b_j^{II}} 
Z(\lambda )
\end{equation}

\begin{equation}
P(a_i^I,b_j^{II}) :: |Z(a_i^I,b_j^{II})|^2
\end{equation}
where $\pi _{A^I} : {\cal P} \to {\cal A}^I$, $\pi _{B^{II }} : {\cal P} \to {\cal B}^{II}$ are the natural projections
from ${\cal P} = {\cal A}^I \times \cdots \times {\cal B}^{II} \times \cdots$ onto the corresponding factors, sets of eigenvalues.

When two systems (e.g., elementary particles) interact some physical magnitudes can become correlated
and some constraints appear in $\cal P$, i.e., equations fulfilled by the eigenvalues, state variables.
The correlation with the wave like component is now expressed as a distribution in the subspace of $\cal P$ determined by the constraints. The simplest example is correlation in one magnitude of each particle, say $A^I + A^{II} = A^T$
with the system in a particular eigenstate of $A^T$, e.g., $|a^T>$ of eigenvalue $a^T$. Magnitudes (eigenvalues) of both particles fulfil $a_i^I + a_i^{II} = a^T$\footnote{A common index $i$ characterizes the correlated pair of eigenvalues.}, and a subset ${\cal P}_{a^T} \subset {\cal P}$ is determined by the constraint.

$Z$ is a distribution of amplitude in ${\cal P}_{a^T}$, equivalently it can be defined to vanish elsewhere.
Consequently, $P(a_i^I,a_j^{II}|$ $a_i^I + a_j^{II}$ $ \neq a^T) = 0$.
Independent measurements of $A^I$ in particle $I$ and $A^{II}$ in particle $II$,
on a correlated pair, has null probability
of giving a result that does not fulfil the correlation. 
The state of the system, $[\lambda _0,Z]$, contains hidden variables $a^I_i =  a$ 
$=\pi _{A^I}(\lambda _0)$, $a^{II}_i =$ $a^T-a$ $= \pi _{A^{II}}(\lambda _0)$,
determining the result of measurements of $A^I$ and $A^{II}$, and marginals

\begin{equation}
Z_i = Z(a^I_i, a^{II}_i) =
\sum _{ \lambda \in {\cal P}_{a^T}; \pi _{A^I}(\lambda ) = a_i^I} Z(\lambda ) 
\end{equation}
determine the observable re\-la\-tive fre\-quen\-cies, 
$P((a^I_i,$ $a^{II}_i)$ $::$ $|Z_i|^2$.
These mar\-gi\-nals define also the orthodox state $|S> = \sum _i$ $Z_i |$ $a^I_i>$ $|a_i^{II}>$. 

We can generalize the previous case to an arbitrary number of correlated magnitudes, to 
obtain a subset of correlated states ${\cal P}_{corr} \subset {\cal P}$, and a distribution of amplitude in
${\cal P}_{corr}$. $A^I$ correlated to $A^{II}$, and $B^{II}$ correlated to $B^I$, are 
now indirectly correlated\footnote{Through the restriction to ${\cal P}_{corr} $ in the sum for marginals.}; $P(a_i^I, b_j^{II})$ is obtained through marginals and Born's rule, as usual. Equivalence with the standard formulation will be established when
the alternative representations
$\sum _i z_i |a^I_i>|a_i^{II}>$, $\sum _j z_j |b^I_j>|b_j^{II}>$,
$\sum _{ij} z_{ij} |a^I_i>|b_j^{II}>$ and 
$\sum _{ji} z_{ji} |b^I_j>|a_i^{II}>$ of the orthodox vector of state are all
obtained as marginals from the common distribution $Z$ in ${\cal P}_{corr}$, 
$|z_i|::|Z_i|$, 
$|z_j|::|Z_j|$, 
$|z_{ij}|::|Z_{ij}|$,  $|z_{ji}|::|Z_{ji}|$,
i.e., modulo normalization and arbitrary phases in the definition of the bases.

\section{Spinless particle}

Relevant magnitudes of an isolated and spinless point particle are position and momentum coordinates;
energy, for example, is redundant $E(p)$. A distribution of amplitude $Z(x,p)$ in the classical phase space, together with 
precise values $(x_0,p_0)$, represents the physical state of the system.
$Z(x,p)$, similarly to Wigner's quasi probability distribution\cite{Wigner} $W(x,p)$, should reproduce the standard representation,
either orthodox amplitudes $\Psi (x)$ and $\xi(p)$, or probabilities 
$P(x)=|\Psi (x)|^2$ and $P(p)=|\xi(p)|^2$ respectively, through marginals.

A simple (not necessarily unique) solution is

\begin{equation}
Z(x,p) = \Psi (x) \xi (p) e^{\frac {i}{\hbar}(x_0 p - x p_0)}
\end{equation}
with $\Psi(x)$ the standard position wave function, and $\xi(p)$ its Fourier transform momentum representation.
Marginals

\begin{equation}
Z(x) = \int dp\,\, Z(x,p) = \Psi (x_0) e^{-\frac {i}{\hbar}x p_0} \Psi (x)
\end{equation}
and 

\begin{equation}
Z(p) = \int dx\,\, Z(x,p) = \xi (p_0) e^{\frac {i}{\hbar}x_0 p} \xi (p)
\end{equation}
are in correspondence with $\Psi(x)$ and $\xi(p)$ respectively,
up to normalization factors\footnote{They do not vanish because the particle is at $(x_0,p_0)$.},
and with the correspondence between bases $\|x> = exp(\frac {i}{\hbar}x p_0)|x>$
and $\|p> = exp(-\frac {i}{\hbar}x_0 p)|p>$, $|x>$ and $|p>$ the standard ones.

The physical interpretation is a point particle with definite position and momentum, although 
both magnitudes can not be jointly and consistently measured and their values must therefore be partially 
hidden\footnote{ Heisenberg's uncertainty principle is fulfilled by the wave functions.}, 
and an accompanying wave, for a wave particle composite whose correlation is represented,
in the ensemble, by the  distribution of amplitude.

In the same way that $\Psi(x)$ is, in the path integral formalism \cite{Feynman}, the integral of elementary amplitudes 
$e^{iS/\hbar}$, $S$ the action integral, over the set of all virtual paths with end point 
$x$, we can 
by analogy interpret $Z(x,p)$ as the result of a path integral over virtual (or abstract) paths with end point $x$ and 
final momentum $p$. 
In the general case,  abstract paths with common value $\lambda$ 
contribute to the amplitude $Z(\lambda )$. 

By linearity, quantum evolution can be expressed as

\begin{equation}
\Psi(x,t_2) =\int dy\,\,K(x,t_2;y,t_1)
\Psi(y,t_1)
\end{equation} 
and it is a matter of 
interpretation to associate the kernel $K(x,t_2;$ $y,t_1)$ to a path integral over all virtual paths joining 
$(y,t_1)$ with $(x,t_2)$. In relativistic mechanics, the integral can be
applied to an arbitrary spatial sheet (in the past or future),
and causality is manifest when restricting the domain of integration;
the value of the amplitude at a given space time event depends
exclusively on values in a spatial sheet inside the past (or future) light cone.
Similarly, in the proposed formalism,

\begin{equation}
Z(x,p,t_2) =\int dy\, dq\,\,K(x,p,t_2; y,q,t_1)Z(y,q,t_1)
\end{equation}
can be easily generalized to the relativistic framework. 
The generic evolution with arbitrary magnitudes is

\begin{equation}
Z(\lambda _2, t_2) = \int d\lambda _1 \,\,K(\lambda _2, t_2;\lambda _1,t_1) 
Z(\lambda _1,t_1)
\end{equation}
for an adequate ``path integral like'' kernel. Evolution of
$\lambda _0(t)$, the corpuscular variables, in a particular system is hidden and stochastic.

\section{The two slit experiment}

In the two slit experiment, a third variable is relevant because of interaction with the slits.
The state of the system is $[(x_0,p_0,S_0), Z(x,p,S)]$, with $x$ and $p$ position and momentum variables at 
the final screen, and $S \in \{L,R\}$ the slit variable, i.e., position at an earlier time.

Marginal amplitude for the final position is

\begin{equation}
Z(x) = \int dp \sum _{S \in \{L,R\}} Z(x,p,S)
 = \int dp \ Z(x,p,L) + \int dp Z(x,p,R) 
\end{equation}
i.e., $Z(x) = Z_L(x) + Z_R(x)$, and gives account of the diffraction pattern, as usual. 
The particle hits the final screen
at position $x_0$, with momentum $p_0$ and coming from slit $S_0$. 
The wave particle interaction,
with wave components coming from both slits, is represented through 
marginals and Born's rule, i.e.,
wave superposition ``guiding'' the particle trajectory, in analogy with Bohm's Mechanics\cite{Bohm}\footnote{But
there is not a deterministic law of evolution for the particle. Wave particle interaction is most probably stochastic.}. In the measurement of the final position,
both momentum $p_0$ and slit $S_0$ are hidden.

Let us suppose an additional system is located at the $R$ slit, and locally interacts with the system, either with
both particle (when going through $R$, $S_0=R$) and wave component at $R$, or
exclusively with the $R$ wave component if the particle goes through $L$, 
$S_0=L$. There will be a change of state,  impulse over the particle 
(for $S_0=R$) and phase shift on the $R$ wave component.
The new marginal is $Z_L(x) + e^{i\theta}Z_R(x)$\footnote{For simplicity, we do not consider a modified $|Z_R|$.}, and we get a displacement of the virtual diffraction pattern.

If the particle is a photon and the additional system an optical plate of fixed phase shift $\theta$, the diffraction pattern is displaced but preserved. If the additional system is
able to show an observable reaction to the presence of the particle,  the phase shift on the wave will be stochastic,
and will appear independently of a positive or negative (indirect measurement) result for the detection of the particle;
the statistical average on $\theta$ destroys the diffraction pattern. 

The result of measurement allows us to distinguish particles arriving to the final screen from either $L$ or $R$ slit,
and observe the relative frequencies $P(x,L)$ and $P(x,R)$,
i.e., $P(x|S_0=L)$ and $P(x|S_0=R)$. Even if both components of the wave are present we can, for all practical purposes, project onto that accompanying  the corpuscular component of the system\footnote{
With known phase shift $\theta$, both wave components interfere for a total distribution
$P(x,\theta) =$ $|Z_L(x) +$ $e^{i\theta}Z_R(x)|^2$.  Photons arriving to $x$ can be (formally) classified according to
the conditional probabilities of coming from the (hidden variable) $L$ or $R$ slit, which are
proportional to $|Z_L(x)|^2$ and  $|e^{i\theta}Z_R(x)|^2$ $=|Z_R(x)|^2$.
Then, $P(x,L,\theta) = P(x,\theta) P(L|x,\theta)$ 
$= |Z_L(x) +$ $e^{i\theta}Z_R(x)|^2$ $\frac {|Z_L(x)|^2}{|Z_L(x)|^2+|Z_R(x)|^2}$,
and $P(x,R,\theta) =$ $|Z_L(x) +$ $e^{i\theta}Z_R(x)|^2$ $\frac {|Z_R(x)|^2}{|Z_L(x)|^2+|Z_R(x)|^2}$.
The average $\int d\theta |Z_L(x) +$ $e^{i\theta}Z_R(x)|^2/(2\pi)$ of the stochastic phase shift under measurement of the slit variable
simplifies  with the denominator, and
the standard projection rule is reproduced.}.
The projection of state appears here as a practical rule, grounded on decoherence between de Broglie wave components.

\section{Bell's experiment}

Let us consider a spin $1/2$ particle. The maximal family of spin variables
contains the spin value in any direction of space; spin operators in two different directions are incompatible, and so no one is redundant. The elementary spin state is a map
$\lambda ({\bf n})$ from
$S^2$ onto $\{+,-\}$, spin up or down, and must be antisymmetric, $\lambda (-{\bf n}) =$ 
$-\lambda ({\bf n})$,
because the spin operators fulfill $S_{-{\bf n}} = - S_{\bf n}$\footnote{In other words, 
the map $\lambda$ acts on the projective space by determining an orientation on each line.}.

We assign a fixed quaternion\footnote{The algebra of quaternions is the Lie algebra 
of rotations.} (instead of complex) amplitude to spin $s$ in direction 
${\bf n}$,
$Z(s, {\bf n}) = s N({\bf n})$, where $N({\bf n})$ is the quaternion $N= n_xI+n_yJ+n_zK$ associated to
the unit vector ${\bf n} = n_x{\bf i}+n_y{\bf j}+n_z{\bf k}$. $I^2=J^2=K^2=-1$,
$I J = - J I = K$, etc., $I^*=-I$, \ldots

We assign to the spin state $\lambda$, all values of spin prescribed, the quaternion amplitude

\begin{equation}
Z(\lambda ) = \frac {1}{4\pi}\int d\Omega \,\,Z(\lambda ({\bf n}),{\bf n})
= \frac {1}{4\pi}\int d\Omega \,\, \lambda ({\bf n}) N({\bf n}) 
\end{equation}
a kind of ``spin path'' integral, sum of elementary amplitudes.

We will restrict next to a finite, but arbitrary, number of directions 
$\{{\bf n}_i\}_{i=1}^n$. 
The former integral becomes

 \begin{equation}
Z(\lambda ) = \sum _i \lambda ({\bf n}_i) N_i = \sum _i s_i N_i
\end{equation}
with $\lambda ({\bf n}_i) = s_i \in \{+,-\}$.  

With this assignment of amplitudes we can now consider
particular ensembles. For example, the quantum state with spin up in direction ${\bf n}_1$
\footnote{Each orthodox spin state for an individual particle is eigenvector of a spin operator.}
is re\-pre\-sented by  the ensemble of elementary states $\lambda _1$ fulfilling 
$\lambda _1({\bf n}_1) = +$, which defines a subset ${\cal P}_{+_1}$ on the phase space of spin states
${\cal P} = \{\lambda \}$.
We either restrict the sum to ${\cal P}_{+_1}$ or declare $Z(\lambda ) = 0$ when 
$\lambda ({\bf n}_1) = -$.

We have marginal $Z(s_1=-)=0$, $Z(s_1=+)\neq 0$, and the orthodox state $|+_1>$ is obtained.
If we calculate marginals $Z(s_2=\pm)$, i.e., the distribution for the basis 
$\{|+_2> ,|-_2>\}$, we get, up to a global factor and using the variables $\lambda \equiv$ $(s_1,s_2,s_3, \ldots, s_N)$,

\begin{equation}
Z(+_2) = \sum _{s_3 s_4 \cdots s_n} Z(+,+,s_3,\ldots s_n) :: +N_1+N_2  
\end{equation}

\begin{equation}
Z(-_2) = \sum _{s_3 s_4 \cdots s_n} Z(+,-,s_3,\ldots s_n) :: +N_1-N_2  
\end{equation}
It is easy to check from the relations $N^*N = 1$ and 

\begin{equation}
N_1^* N_2 = {\bf n}_1\cdot{\bf n}_2 - 
N({\bf n}_1 \times {\bf n}_2) = {\bf n}_1\cdot{\bf n}_2 - ({\bf n}_1 \times {\bf n}_2)_xI -
({\bf n}_1 \times {\bf n}_2)_yJ -({\bf n}_1 \times {\bf n}_2)_zK
\end{equation}
that $|N_1\pm N_2|^2::$ $(1\pm {\bf n}_1\cdot{\bf n}_2)^2$. The orthodox quantum state
in an arbitrary basis is reproduced; in other words, marginals from the global distribution of amplitude match the change of bases in the standard Hilbert space, up to normalization and irrelevant phases in the definition of eigenvectors.

$Z(\lambda _1)$ is the distribution for the ensemble of states with spin up in direction 
${\bf n}_1$.  An individual state has hidden variables $\lambda _1^0$ determining the 
result of a measurement of spin in arbitrary direction. Obviously 
$\lambda _1^0({\bf n}_1) = +$, and $P(\pm_k)$ is found from marginals and Born's rule,
$P(\pm_k) ::|N_1\pm N_k|^2$. Recall that in this interpretation the particle is accompanied by a (we can say here spin) wave, and wave like interference can not be reproduced by an hypothetical classical probability distribution.

Let us consider two particles in a total null spin state, $S_{\bf n}^I + S_{\bf n}^ {II}$
$= S_{\bf n}^T$ ($\forall {\bf n}$) with eigenvalue(s) $0$. In the  space of
elementary spin states $\{\lambda ^T = (\lambda ^I, \lambda ^{II})\}$, the 
constraint $\lambda ^I + \lambda ^{II} = 0$ determines the subset of allowed
states ${\cal P}_0 = \{\lambda ^T_0\}$,
$\lambda ^T_0 = (\lambda ^I, -\lambda ^I)$. The distribution of amplitude 
for the ensemble is defined in 
${\cal P}_0$. The definition $Z^T(\lambda ^T_0) =$ 
$Z(\lambda ^I) - Z(\lambda ^{II})$,
where $Z(\lambda )$ is the former (fixed) quaternion amplitude,
and $Z^T$ is restricted to ${\cal P}_0$, reproduces the quantum results\footnote{$Z^T(\lambda ^I,-\lambda^I) =
Z(\lambda^I)-Z(-\lambda^I)=
2Z(\lambda^I)$.}.
By definition, marginals $Z(s_k^I,s_k^{II})$ with $s_k^I = s_k^{II}$ vanish for all directions ${\bf n}_k$, so that prefect (anti)correlation exists in each correlated pair.

For two different directions, $P(s_1^I, s_2^{II})$, probabilities for up/down results in directions ${\bf n}_1$ for particle $I$ and ${\bf n}_2$ for particle $II$, are calculated in two steps. Marginals 

\begin{equation}
Z(s_1^I,s_2^{II}) = \sum _{\pi _1(\lambda ^I)=s_1^I}\,\,
\sum _{\pi _2(\lambda ^{II}=-\lambda ^I)=s_2^{II}}
\,\, Z^T(\lambda ^I, -\lambda ^{I})
\end{equation}
simplify to

\begin{equation}
Z(s_1^I,s_2^{II}) = {\cal N} (s_1^I N_1 - s_2^{II}N_2)
\end{equation}
$\cal N$ a global factor. Relative frequencies are found trough Born's rule

\begin{equation}
P(s_1^I, s_2^{II}) :: |s_1^I N_1 - s_2^{II}N_2|^2 = 
2(1 - s_1^I\, s_2^{II}\, {\bf n}_1 \cdot {\bf n}_2)
\end{equation}
Recall that hidden variables $\lambda ^T_0 \in {\cal P}_0$ accompany the distribution $Z^T$ for a complete description of an individual state of the composite. The result of  measurements in arbitrary directions on each particle of a correlated
pair is prescribed from the generation event.

It is interesting to see explicitly the analogy between interference behavior
in the two slit experiment and interference like behavior of the ``spin'' wave.
If we consider another direction ${\bf n}_3$ of hypothetical measurement of particle $II$, the marginals are up to global factor

\begin{equation}
Z(s_1^I,s_2^{II},s_3^{II}) = s_1^IN_1 -(s_2^{II}N_2 + s_3^{II}N_3)
\end{equation}
We get from there a formal distribution of probability
$P(s_1^I,s_2^{II},s_3^{II})$ $::|s_1^IN_1 -$ $(s_2^{II}N_2 + s_3^{II}N_3)|^2$, but $P(s_1^I, s_2^{II})$ is not the marginal $\sum _{s_3^{II}=\pm}$
$P(s_1^I,s_2^{II},s_3^{II})$, because
there are ``interference'' ($[[\ldots]]$) terms

\[
|Z(s_1^I,s_2^{II},+_3^{II}) + Z(s_1^I,s_2^{II},-_3^{II})|^2 =
\]

\[
|s_1^I N_1 - (s_2^{II} N_2 + N_3)|^2 + 
|s_1^I N_1 - (s_2^{II} N_2 - N_3)|^2 +
\]

\[
[[Z^*(s_1^I,s_2^{II}, +_3^{II})Z(s_1^I,s_2^{II},-_3^{II}) +
Z^*(s_1^I,s_2^{II}, - _3^{II})Z(s_1^I,s_2^{II},+_3^{II})]]
\]
as in the two slit experiment, where we have

\[
|Z_L(x)+Z_R(x)|^2 =|Z_L(x)|^2+ |Z_R(x)|^2 +
[[Z^*_L(x)Z_R(x) + Z^*_R(x)Z_L(x)]]
\]

The analogy can be extended to Bell's theorems,
if we consider the following trivial inequality: two strictly positive
distributions of probability for independent events can not generate a marginal distribution with zeros.
If we  consider a point particle arriving to the final screen from either $R$ or $L$ slit as an isolated system, particles through $R$ and particles through $L$ are independent events. Then,

\begin{equation}
\sum _{\lambda}P(x,L,\lambda) + 
\sum _{\lambda '}P(x,R,\lambda') \geq
{\rm max}\{\sum _{\lambda}P(x,L,\lambda),\sum _{\lambda '}P(x,R,\lambda')\}> 0
\end{equation} 
with $\lambda$, $\lambda'$ additional variables, e.g., momentum. The diffraction pattern can not be reproduced.
An additional wave like system, accompanying the point particle, 
can give account of the phenomenon.

\section{Physical interpretation}

Standard QM contains the maximal amount of information we can have about a quantum system. In an interaction, magnitudes not commuting with the Hamiltonian of interaction can change of value. Measurement is an interaction. Therefore, only a maximal family of commuting magnitudes can be jointly and consistently measured. However, it seems to be incomplete:

\begin{enumerate}
\item The spatial region where the wave function has significant values is much larger than the observed region where the corpuscular system is located.

\item Measurement, the projection of state, violates the principle of local interactions.

\item The action reaction principle, symmetric
effects over both systems in interaction, is also violated.
\end{enumerate}

An extended phase space contains hidden variables, unknown values of some magnitudes of the system; measurement of any of these variables unavoidably destroys the information about previously known precise values of other variables of state. Hypothetical distributions of probability in the extended phase space are not observable.
Moreover, Bell's theorems forbid the existence of distributions of probability whose marginals match the quantum predictions. Points, elements of this phase space, can not be a complete description of an individual system of the ensemble.

In the proposed formulation of QM, a complex distribution of amplitude is the re\-pre\-sentation of an ensemble of quantum systems, each individual system being a composite of corpuscular and wave like subsystems. 
Variable of state $\lambda _0$ of the corpuscular component(s) prescribes for each indivi\-dual system of the ensemble the
result of an arbitrary direct measurement, but it is not a complete description of the composite. Marginal amplitudes and Born's rule 
determine observable relative frequencies in the ensemble. The interference that appears in the sum of amplitudes generates wave like behaviour, and it is interpreted as a statistical representation of the correlation between the de Broglie wave and corpuscular component in the composite.

Isolated effects of a wave component in indirect measurements,
when the particle is spatially separated from the apparatus, are predicted in this formulation in order to preserve both locality and the ARP. These effects are absent in standard QM, where only results and probabilities of direct measurements are described. 
Such effects have not been observed; equivalence of predictions with the orthodox theory for direct measurements makes this formalism an alternative interpretation of QM.

The physical character of the de Broglie waves is a deeper issue. Vacuum is considered a relevant physical system,
which stores energy and interacts with matter and radiation, at both ends of the length scale. In Quantum Field Theory, it is source of annoying divergences, and of the Casimir effect. In Cosmology, dark (vacuum) energy density  is responsible of the observed accelerated expansion of the universe. At intermediate scales, vacuum energy density could be behind dark matter, if it were not homogeneously distributed. All we can say about dark matter is it is dark and stores energy,
has gravitational pull, which fits well with the vacuum hypothesis.
Vacuum fluctuations could be the physical interpretation of the de Broglie waves, as in Quantum Field Theory. 
The de Broglie wave accompanies 
all particles and quantum systems, i.e., it must be independent of known fields as, e.g., the electromagnetic.  It must be associated to an \ae ther.

There is a formal \ae ther, a Lorentz invariant one. It is the volume form in the light cone of momentum space at each point of space time, understood as a distribution of density of particles with null rest mass. Total volume of the cone is obviously infinite (both density of particles and energy), and represents a classical counterpart of quantum divergences.
Total momentum vanishes by symmetry, in all inertial frames!

Local, spatially inhomogeneous, displacements from this equilibrium distribution store (positive or negative) finite energy increment, as well as non vanishing momentum. In a classical fluid there are relevant wave like phe\-no\-me\-na at all length scales, from Brownian motion to Earth scale tides.
It is appealing to consider a parallelism with quantum fluctuations and dark energy. The classical \ae ther hypothesis (rejected because there is not a distinguished rest frame) has a relativistic invariant and divergent counterpart, which could be relevant too in the theory of gravitation.

\section{Acknowledgements}

Financial support from research project MAT2011-22719 is acknow\-ledged. I
kindly ack\-now\-led\-ge helpful comments from members of the audience in both
seminars at Zaragoza and Valladolid Universities, where I presented some
conclusions of this research in April and June 2015 respectively.


\begin{thebibliography}{}






\bibitem{Aspect} A. Aspect, P. Grangier and G. Roger : {\em Phys. Rev. Lett.}
{\bf 49} 1804  (1982)



\bibitem{Bell} J. S. Bell :  {\em Physics 1}, 195 (1964)

\bibitem{Bohm} D. Bohm : {\em Phys. Rev.} {\bf 85} 643  (1952)


\bibitem{Born} M. Born : {\em Z. Phys.} {\bf 37}  863 (1926)

\bibitem{deBroglie} L. de Broglie : {\em Annales de la Fondation Louis de Broglie}
{\bf 12}(4) 1 (1987)


\bibitem{EPR} A. Einstein, B. Podolsky and N. Rosen : {\em Phys. Rev.}{\bf 47},
777 (1935)

\bibitem{Feynman} R. P. Feynman and A. R. Hibbs : {\em QM and Path Integrals}
McGraw--Hill, New York (1965)





\bibitem{mio} C. L\'opez : {\it A local interpretation of QM},
arXiv:1412.5612. Sent to {\em Foundations of Physics}




\bibitem{Wigner} E. P. Wigner : {\em Phys. Rev. }{\bf 40} 749 (1932)








\end{thebibliography}
\end{document}